\documentclass[a4paper,accepted=2024-05-17]{quantumarticle}
\pdfoutput=1

\usepackage[numbers]{natbib}
\usepackage{amsmath}
\usepackage{amssymb}
\usepackage{graphicx}
\usepackage{dcolumn}
\usepackage{bm}
\usepackage{multirow}
\usepackage{xcolor}

\newtheorem{proposition}{Proposition}
\usepackage{physics}
\usepackage{hyperref}
\usepackage{cleveref}

\begin{document}


\title{Empirical Sample Complexity of Neural Network Mixed State Reconstruction}

\author{Haimeng Zhao}
\affiliation{Institute of Physics, \'{E}cole Polytechnique F\'{e}d\'{e}rale de Lausanne (EPFL), CH-1015 Lausanne, Switzerland}
\affiliation{Center for Quantum Science and Engineering, \'{E}cole Polytechnique F\'{e}d\'{e}rale de Lausanne (EPFL), CH-1015 Lausanne, Switzerland}
\affiliation{Zhili College, Tsinghua University, Beijing 100084, China}

\author{Giuseppe Carleo}
\affiliation{Institute of Physics, \'{E}cole Polytechnique F\'{e}d\'{e}rale de Lausanne (EPFL), CH-1015 Lausanne, Switzerland}
\affiliation{Center for Quantum Science and Engineering, \'{E}cole Polytechnique F\'{e}d\'{e}rale de Lausanne (EPFL), CH-1015 Lausanne, Switzerland}

\author{Filippo Vicentini}
\affiliation{Institute of Physics, \'{E}cole Polytechnique F\'{e}d\'{e}rale de Lausanne (EPFL), CH-1015 Lausanne, Switzerland}
\affiliation{Center for Quantum Science and Engineering, \'{E}cole Polytechnique F\'{e}d\'{e}rale de Lausanne (EPFL), CH-1015 Lausanne, Switzerland}
\affiliation{CPHT, CNRS, \'{E}cole polytechnique, Institut Polytechnique de Paris, 91120 Palaiseau, France}
\affiliation{Coll\`ege de France, Universit\'e PSL, 11 place Marcelin Berthelot, 75005 Paris, France}


\begin{abstract}
    Quantum state reconstruction using Neural Quantum States has been proposed as a viable tool to reduce quantum shot complexity in practical applications, and its advantage over competing techniques has been shown in numerical experiments focusing mainly on the noiseless case.
    In this work, we numerically investigate the performance of different quantum state reconstruction techniques for mixed states: the finite-temperature Ising model.
    We show how to systematically reduce the quantum resource requirement of the algorithms by applying variance reduction techniques. 
    Then, we compare the two leading neural quantum state encodings of the state, namely, the Neural Density Operator and the positive operator-valued measurement representation, and illustrate their different performance as the mixedness of the target state varies.
    We find that certain encodings are more efficient in different regimes of mixedness and point out the need for designing more efficient encodings in terms of both classical and quantum resources.
\end{abstract}

\maketitle

\section{Introduction}
Recent advances in quantum technologies \cite{nielsen2010quantum} have led to diversified applications in areas such as quantum simulation \cite{georgescu2014quantum}, communication \cite{gisin2007quantum}, cryptography \cite{gisin2002quantum} and machine learning \cite{biamonte2017quantum}.
However, present-day \textit{noisy intermediate-scale quantum} (NISQ) devices are inherently noisy \cite{preskill2018quantum} and limited in size. 
Techniques to mitigate noise \cite{cai2022quantum},  reduce quantum circuit depth \cite{mcclean2016theory}, minimize the number of quantum shots, and verify experimentally prepared states \cite{carrasco2021theoretical} are crucial to leveraging these devices in realistic applications.

One such technique, quantum state reconstruction (or tomography) \cite{paris2004quantum}, uses a limited number of measurements to produce a classical approximation $\hat{\rho}$ of the quantum state $\rho$ prepared on a device.
The \textit{classical reconstruction} facilitates the efficient computation of many observables without further quantum circuit evaluations and serves to validate the quantum device \cite{carrasco2021theoretical}.
Classical reconstructions, which vary in computational overhead, are all derived from datasets of measurement outcomes collected by repeatedly measuring a set of observables on the state $\rho$ prepared on the device.
The accuracy of the reconstruction can be quantified via several \textit{reconstruction errors} $\epsilon$, such as the difference in expectation values between the reconstruction and the original state or other distance measures like infidelity or trace distance between $\hat{\rho}$ and $\rho$.
An important indicator of the asymptotic performance of such methods is the sample complexity: the size of the \textit{quantum-generated dataset} needed to obtain a classical reconstruction with a certain error $\epsilon$.

Recent research has established that tomography methods using generic quantum states, such as maximum likelihood estimation \cite{lvovsky2004iterative, vrehavcek2001iterative}), necessitate a sample size that grows exponentially with the system size \cite{flammia2012quantum, haah2016sample, yuen2023improved}.
One way to circumvent this is to design randomized measurement protocols that only estimate expectation values of certain observables (e.g., classical shadow tomography \cite{aaronson2018shadow, huang2020predicting, hadfield2022measurements, elben2023randomized})
Alternatively, one can exploit the fact that only a small set of possible states, with low complexity, are actually observed in physical models \cite{poulin2011quantum, brandao2021models, zhao2023learning}, and design more efficient encodings of those physically realizable states to alleviate the data requirement, similar to variational methods commonly adopted to simulate quantum systems classically. 
For example, one can use matrix product states to efficiently encode and reconstruct one-dimensional pure states with area law entanglement \cite{baumgratz2013scalable, lanyon2017efficient, guo2024scalable}.
For more general states, generative neural networks (NN) can be used as a variational ansatz trained to reproduce the measurement data \cite{carleo2017solving, torlai2018neural, carrasquilla2019reconstructing, lohani2021experimental, schmale2022efficient, cha2021attention, torlai2018latent}.
These ansatze, called Neural Quantum States (NQS), encode the quantum state in a compact form, significantly reducing the data requirement for reconstructions. The cost associated with this approach is introducing a non-convex optimization problem, which established techniques in modern artificial intelligence can approximately solve \cite{goodfellow2016deep, torlai2020precise}.

Designing efficient NQS encodings for general mixed states presents a significant challenge. 
Three broad categories exist: (i) an expansion onto a set of pure-states encoded with standard NQS, which is exponentially costly for states with large entropy \cite{Melkani2020PRAEigenstateExtraction}; (ii) a physical (positive-semidefinite) neural-network encoding of the mixed state in the Pauli-Z computational basis known as Neural Density Operator (NDO) \cite{torlai2018latent, vicentini2022positive} and (iii) a neural network encoding of the probability distribution over the outcomes of a set of informationally-complete positive operator-valued measurements (POVM-NQS).
This last approach is less costly than (ii), but may result in unphysical density matrices \cite{carrasquilla2019reconstructing, cha2021attention, schmale2022efficient, Neugebauer20PRAPovmVsNQS}.
While the approximation used in the first approach is well-understood, the relationship between the latter two methods remains unclear. 
Furthermore, no comparison has been made thus far regarding their effectiveness in Quantum State Reconstruction tasks.
In particular, whether these two methods share the same sample complexity is uncertain.

While the dependence on system size has been extensively studied for pure states (e.g., \cite{sehayek2019learnability} Fig. 7), few comparisons have been made regarding the dependence on reconstruction error $\epsilon$. 
This $\epsilon$ dependence can be especially significant for NISQ algorithms such as variational quantum eigensolvers (VQEs) \cite{kandala2017hardware, mcclean2016theory}, as a worse scaling will lead to a large increase in the number of quantum executions to achieve the same accuracy of the result.
For classical shadow tomography, the sample complexity is known to scale as $\epsilon^{-2}$, which doesn't yield an asymptotic improvement over naive statistical averaging \cite{huang2020predicting, lukens2021bayesian}.
More recently, numerical evidence suggests that the pure state NQS method holds an approximately quadratic advantage (i.e., $\epsilon^{-1}$) over classical shadow in energy estimation for certain molecular ground states \cite{iouchtchenko2023neural}.
However, it is unknown if this advantage persists when the target state is mixed.

In this work, we first improve the reconstruction algorithm by considerably reducing its classical computational overhead using the Control-Variates variance reduction technique (\cref{sec:cv}).

Then, we conduct comprehensive numerical simulations to investigate the sample complexity of mixed-state reconstruction for different NQS encodings.
In particular, we benchmark NDO and POVM-NQS on reconstructing the finite temperature density matrix of the Transvese-Field Ising model.
We numerically demonstrate that for NDO, the quadratic advantage in pure state reconstruction can only survive when the state is slightly mixed, and the scaling deteriorates when the state is highly mixed.
On the other hand, POVM-NQS does not hold such an advantage even for pure states and has a similar scaling as classical shadows, independent of how mixed the target state is. 
Consequently, NDO performs better than POVM-NQS for nearly pure states, while for highly mixed states, the situation is reversed.
We also propose a phenomenological model that can explain the results.
These results provide valuable guidance to the practical implementation of NQS-based state reconstruction and also point out the need for designing more efficient encodings in terms of quantum resources.

\section{Neural Quantum State Reconstruction}
We begin by describing the general framework of NQS reconstruction. 
The fundamental concept involves training a (potentially generative) NN that approximates a quantum state in a well-defined basis to reproduce the statistics of the measurement data. 
Let's assume that in the experiments, the system state $\rho$ has been measured with $N_b$ different POVM measurements $\{\{P^b_i\}_{i=1}^K\}_{b=1}^{N_b}$, where $\sum_{i=1}^K P^b_i=I$. 
This includes projective measurements like Pauli string measurements as a special case. 
Assume that we have gathered $N_d$ measurement outcomes under each measurement basis, where each outcome is denoted by a number $\sigma^b_j \in \{1, \ldots, K\}$ for $j = 1, \ldots, N_d$. 
The dataset that we aim to reproduce can be represented as: 
\begin{equation}
    D = \bigcup_{b=1}^{N_b} D_b = \bigcup_{b=1}^{N_b} \{\sigma^b_j\}_{j=1}^{N_d}.
\end{equation}
 
We use $p_b(\sigma)$ to denote the probability of obtaining the measurement outcome $\sigma$ by measuring $\rho$ under basis $b$, and use $q^\theta_b(\sigma)$ to denote the corresponding probability given by the NQS $\rho_\theta$ with variational parameters $\theta$. 
Our goal is to minimize the averaged distance between these two probability distributions over all bases.
We quantify this distance by the Kullback–Leibler divergence $\mathrm{KL}(p\|q)=\mathbb{E}_{\sigma\sim p}\log \frac{p(\sigma)}{q(\sigma)}$. 
We, therefore, define the loss function as:
\begin{equation}
\label{eq:loss}
\begin{split}
    \mathcal{L}(\theta) &= \frac{1}{N_b} \sum_{b=1}^{N_b} \text{KL}(p_b \| q^\theta_b) \\
    &\approx \frac{1}{N_b} \sum_{b=1}^{N_b} \frac{1}{N_d} \sum_{j=1}^{N_d} \log q_b^\theta(\sigma^b_j) + \mathrm{const.},
\end{split}
\end{equation}
where we have omitted the constants that do not depend on $\theta$ and approximated the expectation values $\mathbb{E}$ with the sample average over the finite dataset. 
We then use gradient-based optimization methods to optimize the parameters $\theta$, in which the $t^\mathrm{th}$ iteration reads
\begin{equation}
    \theta_t = \mathrm{opt}(\theta_{t-1}, \nabla_\theta \mathcal{L}(\theta_{t-1})),
\end{equation}
where $\mathrm{opt}$ refers to the optimization algorithm used (e.g, for gradient descent with learning rate $\alpha$, $\mathrm{opt}(\theta, g) = \theta - \alpha g$). 
In practice, when the dataset is large, we employ a technique called mini-batching to reduce the computational cost.
This involves estimating the KL-divergence and its gradient not on the whole dataset, but only on a smaller subset of it. 
Once the training procedure has converged, the NQS can be used to generate samples, predict properties of interest, or, for sufficiently small systems, retrieve the full density matrix. 

We now briefly introduce the two different NQS encodings that we compared during our investigations: NDO and POVM-NQS.

\subsection{Neural Density Operator}
The NDO encoding is compatible with projective measurements. 
We take Pauli string measurements $\{X, Y, Z\}^{\otimes N}$ on an $N$-qubit system as an example. 
The corresponding projectors will be denoted with $\{P^b_i\}$ and the basis rotation matrix with $\{U_b\}$. 
Measurement outcomes can be denoted by bit-strings of length $N$ (e.g.,  $\sigma^b_j = (0100)$ for $N=4$). 
When a properly normalized neural network (NN) is used - for instance, an autoregressive NN - to parameterize the density matrix elements $\matrixelement{\eta}{\rho_\theta}{\eta'} = \mathrm{NN}_\theta(\eta, \eta')$, the variational probability distribution reads
\begin{equation}
\label{eq:ndo}
    q^\theta_b(\sigma) = \sum_{\eta, \eta'}\matrixelement{\sigma}{U_b}{\eta}\matrixelement{\sigma}{U_b}{\eta'}^*\mathrm{NN}_\theta(\eta, \eta').
\end{equation}
If the NN parameterization is not normalized, an additional normalization term in the loss function appears, whose gradient can be estimated through Monte Carlo sampling \cite{torlai2018neural}. 
We remark that to efficiently evaluate \cref{eq:ndo}, the number of connected elements in the rotation unitary $U_b$ must be bounded.
However, their number will grow exponentially with the number of $X$s and $Y$s rotations present in the measurement basis.
A similar limitation appears for pure-state tomography.

For benchmarking purposes, we use all the possible $3^N$ basis in this work, to guarantee that the dataset contains enough information to determine the mixed target state.
Nevertheless, this is not necessary in practice since NDO works for generic datasets and measurement basis that may not be informationally complete \cite{torlai2018latent,sehayek2019learnability,iouchtchenko2023neural}.

\subsection{POVM-NQS}
The POVM-NQS method exploits the one-to-one correspondence between the state $\rho$ and the outcome statistics of a single ($N_b=1$) informationally complete POVM measurement $\{P_i\}_{i=1}^K$.
The probability of obtaining an outcome $\sigma \in \{1, \ldots, K\}$ is given by $p(\sigma) = \tr(\rho P_\sigma)$. 
Conversely, the density matrix can be obtained by the inverse formula 
\begin{equation}
    \rho = \sum_{\sigma, \sigma'}p(\sigma') T^{-1}_{\sigma \sigma'} P_{\sigma},
\end{equation}
where $T_{\sigma, \sigma'} = \tr(P_{\sigma} P_{\sigma'})$ is the overlap matrix \cite{carrasquilla2021probabilistic}. 
Therefore, reconstructing $p(\sigma)$ suffices to determine $\rho$.
As an example, we consider the tensor products of single-qubit Pauli-4 measurements $\{P_{(0), (1), (2)} = \frac{1}{3}\ket{\uparrow_{x, y, z}}\bra{\uparrow_{x, y, z}}, P_{(3)}=I - P_{(0)} - P_{(1)} - P_{(2)} \}^{\otimes N}$ for an $N$ qubit system. 
Under this measurement scheme,  a normalized neural network is used to approximate $p(\sigma)$: $q^\theta(\sigma) = \mathrm{NN}_\theta(\sigma)$. 
When the NN is trained, one can use the inverse formula to reconstruct the target state, or directly estimate relevant properties via sampling \cite{schmale2022efficient}.

\section{Variance Reduction via Control Variates}
\label{sec:cv}

\begin{figure}
    \centering
    \includegraphics[width=0.9\linewidth]{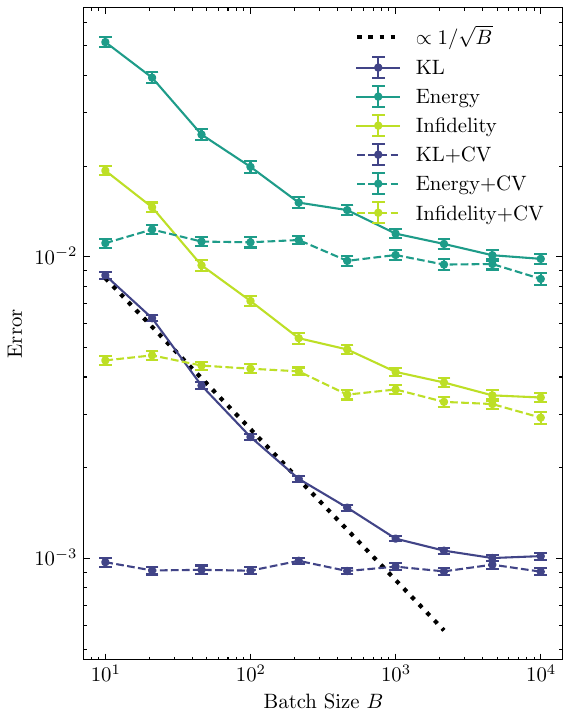}
    \caption{NQS reconstruction of the ground state of 3-qubit 1D open-boundary transverse field Ising model using NDO with different batch sizes $B$. 
    KL divergence, energy error and infidelity are used as metrics for performance. Dashed and solid lines represent the results of training with and without the control variates method. The dotted line marks the $1/\sqrt{B}$ scaling. All the data points are averaged over 100 random instances.}
    \label{fig:control_variate}
\end{figure}

To investigate the asymptotic behavior of the sample complexity, we must train the NN to high precision. 
However, we observe that the noise introduced by the mini-batching strategy makes accurate training prohibitively expensive in practice. 
To understand that, consider randomly sampling a batch of outcomes $B$ from the dataset at each iteration. 
The gradient of the loss function evaluated on this batch $\mathcal{L}_B$ will be given by
\begin{equation}
    g_B(\theta) = \nabla_\theta \mathcal{L}_B = \frac{1}{|B|}\sum_{\sigma^b_j\in B} \nabla_\theta \log q^\theta_b (\sigma^b_j),
\end{equation}
which an unbiased estimator of $\nabla_\theta \mathcal{L}$. 
However, the variance of $\mathcal{L}_B$ reads
\begin{equation}
    \label{eq:variance}
    \mathrm{Var}[\mathcal{L}_B] = \frac{1}{|B|} \mathrm{Var}_{\{\sigma^b\sim p_b\}}\left[\frac{1}{N_b} \sum_{b=1}^{N_b} \log q^\theta_b(\sigma^b)\right],
\end{equation}
which remains finite as $q^\theta_b$ approaches the target $p_b$. 
This asymptotically finite variance is in sharp contrast to the zero-variance property of Variational Monte Carlo, which allows for accurate optimization of the ground-state with a relatively small number of samples \cite{becca2017quantum, sinibaldi2023unbiasing}.
The statistical fluctuations in the gradient estimation introduce noise that prevents the reconstruction error from dropping below its standard deviation, which scales proportionally to $1/\sqrt{|B|}$. 
Consequently, in situations where we cannot afford training with large batch sizes, we must reduce the variance of the gradient estimator.

\begin{figure*}
    \centering
    \includegraphics[width=\linewidth]{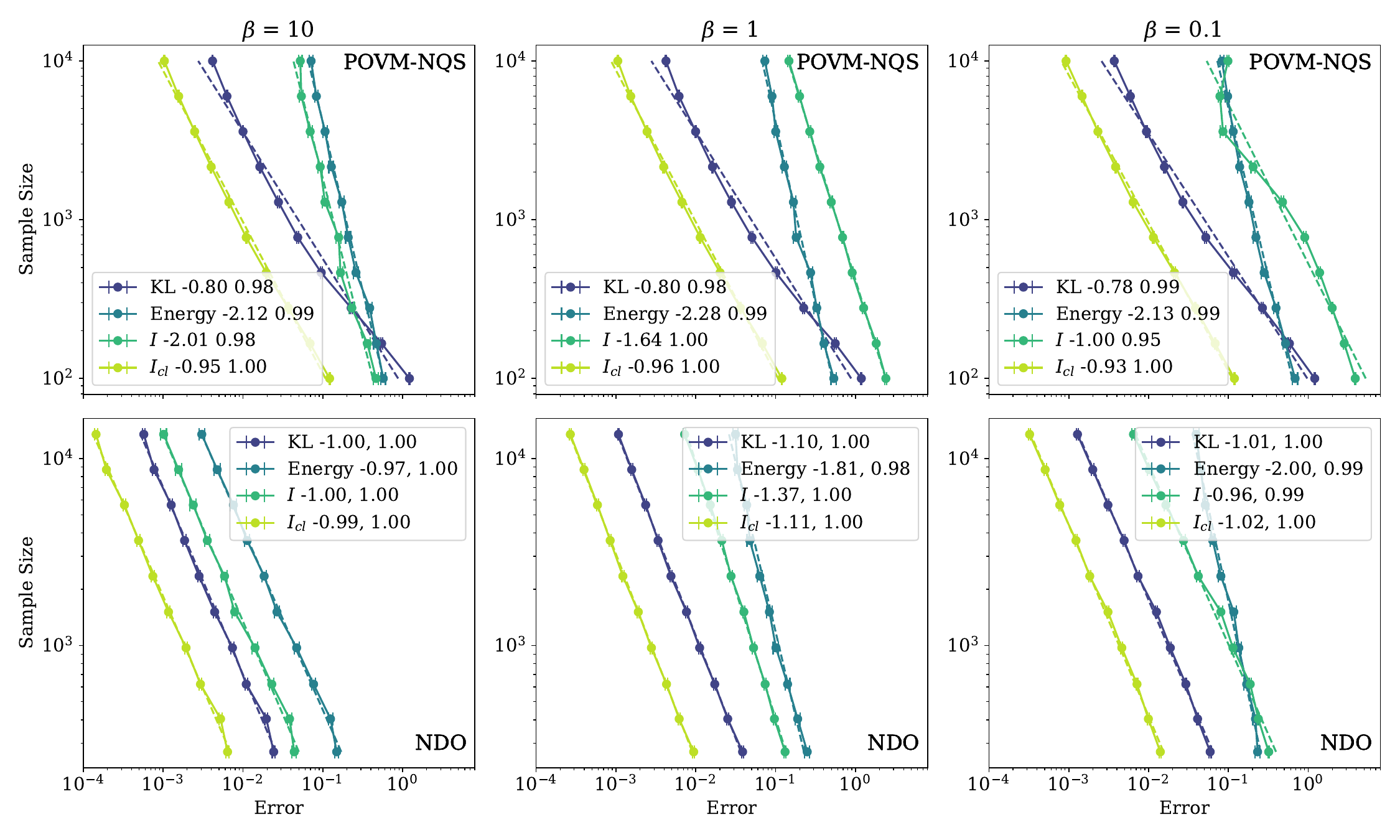}
    \caption{Sample complexity dependence on the reconstruction error for 3-qubit 1D open-boundary transverse field Ising model under different inverse-temperature $\beta=10, 1, 0.1$ using POVM-NQS (top) and NDO (bottom). KL divergence, energy error, infidelity and classical infidelity are used as metrics for reconstruction error. Dashed lines represent the log-log linear regression results, with the slopes and $r^2$ values indicated in the legend. All data points are averaged over 100 random instances.}
    \label{fig:scaling_ising}
\end{figure*}

The Control Variates (\textbf{CV}) method is a well-established statistical technique for variance reduction \cite{boyle1977options}, which was recently discussed in the context of Variational Monte Carlo \cite{wu_solving_2019, sharir_deep_2020, hibat2020recurrent, sinibaldi2023unbiasing}.
Specifically, when estimating the gradient $g_B(\theta_t)$ at step $t$, we introduce a second random variable (the CV), which represents the gradient evaluated at an earlier step $t'$: $g_B(\theta_{t'})$. 
We then adjust the gradient estimator to $g_B(\theta_{t-1}) - g_B(\theta_{t'}) + \mathbb{E}g_B(\theta_{t'})$.
This revised estimator is unbiased but can have lower variance because $g_B(\theta_t)$ and $g_B(\theta_{t'})$ are correlated.
The expectation can be computed by averaging over the entire dataset. 
Consequently, we obtain the following variance-reduced training rule:
\begin{equation}
    \theta_t = \mathrm{opt} (\theta_{t-1}, g_B(\theta_{t-1}) - g_B(\theta_{t'}) + \nabla_\theta\mathcal{L}(\theta_{t'})).
\end{equation}
To further reduce the computational cost, we update the CV only once every $T$ steps, i.e., $t' = T \lfloor t/T \rfloor$. 
For all simulations in this paper, we set $T=50$.
This method is also known as \textit{stochastic variance reduced gradient} (SVRG) in the machine learning literature \cite{johnson2013accelerating, mohamed2020monte}. 

To further substantiate our approach, we conduct a systematic numerical analysis.
As a benchmark problem, we consider the NDO reconstruction of the one-dimensional open-boundary transverse field Ising model (TFIM)
\begin{equation}
    H_{\mathrm{Ising}} = -\sum_{i=1}^{N-1}Z_iZ_{i+1} - h\sum_{i=1}^N X_i.
\end{equation}
We set $h=1$ and $N=3$, small enough to study the batch size's effects systematically. 
We randomly generate $10^3$ measurement shots for each of the $3^3=27$ Pauli basis, train the NDO with and without CV for different values of the batch size $B$, and repeat every simulation 100 times with different initial conditions and random seeds. 
The details of the training process and NDO are listed in \cref{app:numeric}.
In \cref{fig:control_variate}, we compare three metrics, the KL divergence averaged over all measurement basis (KL), the error in energy $\varepsilon = |\tr(H\rho_\theta)-\tr(H\rho)|$, and the infidelity $I(\rho, \rho_\theta) = 1-\left(\tr\sqrt{\sqrt{\rho_\theta}\rho\sqrt{\rho_\theta}}\right)^2.$ 
Dashed and solid lines correspond to training with and without CV. 
It is evident that when NQS is trained without CV, the errors scale like $1/\sqrt{B}$ as expected from \cref{eq:variance}, and then saturates for large batch size at an intrinsic limit set by the dataset. 
When the CV method is applied, the adverse effect of mini-batching is eliminated, and the errors are independent of the batch size.
These results validate the effectiveness of our CV method, which we will use in the rest of our analysis.

We note that this CV method applies to generic NQS reconstruction algorithms, including different pure and mixed encodings, and should be used as a default technique when mini-batching introduces noise.
The code is implemented and open-sourced in NetKet \cite{netket2:2019, netket3:2022, mpi4jax:2021}.

\section{Results and Discussion}

\begin{figure}
    \centering
    \includegraphics[width=\linewidth]{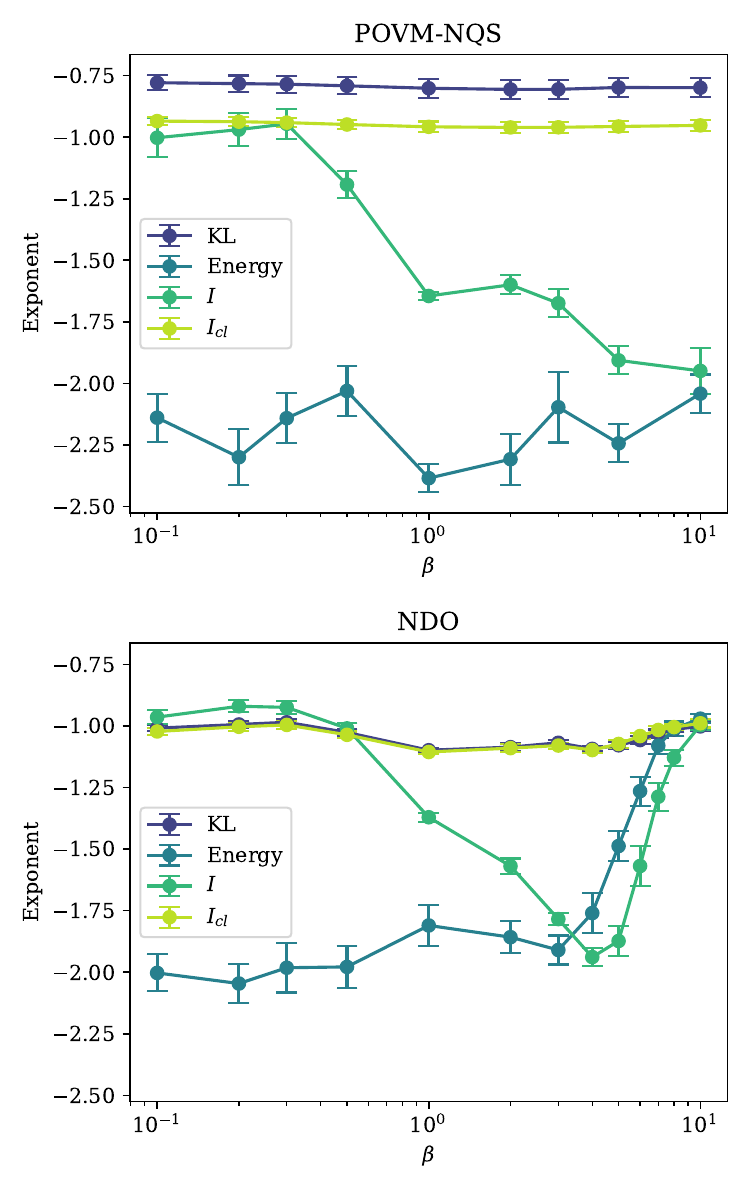}
    \caption{The scaling exponents of reconstruction error for 3-qubit 1D open-boundary transverse field Ising model under different inverse-temperature $\beta \in [10^{-1}, 10^1]$ with POVM-NQS (top) and NDO (bottom). KL divergence, energy error, infidelity, and classical infidelity are used as metrics for reconstruction error. The error bars are given by linear regression.}
    \label{fig:exp_ising}
\end{figure}

\subsection{Simulations}

To study the performance of NDO and POVM-NQS, we simulate the finite-temperature Gibbs ensemble of the TFIM, which is representative of mixed states where the prepared states might interact with a thermal bath.

We generate measurement datasets of varying sizes and use different NQS ansatze for the reconstruction to understand the asymptotic scaling behavior of the sample complexity. 
In this work, we focus on sample sizes in the regime of $10^2$ to $10^4$, which are currently achievable with modern quantum devices \cite{arute2019quantum, endres2016atom}.
By plotting the errors of different sample sizes on a log-log scale, we can determine the sample complexity scaling exponents from the slopes of the linear fits.
We use the loss value (defined in \cref{eq:loss}), the infidelity $I$, and the error in energy $\varepsilon$ as metrics for the reconstruction error. 
As the density matrix reconstructed through the POVM-NQS method might be negative, the infidelity may not always be a good indicator.
Therefore, we take the absolute value of infidelity and also calculate the average classical infidelity $I_{cl} = 1 - \frac{1}{N_b}\sum_{b}\sum_{\sigma}\sqrt{p_b(\sigma)q_b^\theta(\sigma)}$, which is commonly used as a performance indicator in the literature on POVM-NQS.

We consider the 3-qubit one-dimensional open-boundary TFIM at $h=1$, and use thermal states $\rho_\beta = \exp(-\beta H_{\mathrm{Ising}})/\tr[\exp(-\beta H_{\mathrm{Ising}})]$ across a wide range of inverse-temperatures $\beta\in[10^{-1},10^1]$ as the target states. 
The numerical details can be found in \cref{app:numeric}.
In Fig. \ref{fig:scaling_ising}, we show three inverse-temperatures $\beta = 10, 1, 0.1$ representing low, medium, and high-temperature regimes, respectively. 
We plot the sample complexity scaling behaviors using solid lines, with the linear fits shown as dashed lines. 
The slopes and r-squared values of the linear fits are reported in the corresponding legends. 
In Fig.~\ref{fig:exp_ising}, we summarize the scaling exponents for different inverse temperatures.

\subsection{Scaling Behavior}

As shown in \cref{fig:exp_ising}, we note that the scaling exponents for KL and classical infidelity for both classes of NQS ansatze are approximately $-1$, regardless of the mixedness of the target states. 
This is because we are learning the classical probability distributions of the measurement outcomes, which is known to have sample complexity $\Theta\left(\epsilon^{-1}\right)$, for a classical error $\epsilon$ quantified by the KL or classical infidelity \cite{canonne2020short}.

Regarding energy, POVM-NQS and NDO show qualitatively different behaviors. 
The scaling exponent for POVM-NQS remains around $-2$ irrespective of the mixedness of the target state, suggesting that the method doesn't improve the asymptotic quantum shot complexity over naive statistical averaging or classical shadows. 
On the other hand, NDO exhibits a scaling exponent of approximately $-1$ for slightly mixed states, although this exponent gradually deteriorates to $-2$ for highly mixed states. 
This suggests that NDO has an advantage over POVM-NQS when the target state is only slightly mixed, but the advantage disappears when the mixedness increases significantly. 
As NDO ansatze can naturally represent pure-states, this observation is consistent with the behavior of pure-state reconstruction, which was recently numerically also demonstrated to have an exponent of $-1$ (see Ref.~\cite{iouchtchenko2023neural}).  
We further note that NDO tends to be more accurate than POVM-NQS in terms of reconstruction error for the same sample size. Still, the classical optimization process is often harder to converge, leading to higher classical overhead.
This might be related to variance problems arising from zeroes in the density matrix, similar to what was recently found for variational dynamics in Ref.~\cite{sinibaldi2023unbiasing}.

In \cref{app:LiH}, we conduct the same study for a molecular ground-state (LiH) subject to depolarization and find a consistent picture as here.

\subsection{Theoretical Analysis}
\label{sec:theory}

\begin{table}
\begingroup
\setlength{\tabcolsep}{6pt} 
\renewcommand{\arraystretch}{1.5} 
\begin{tabular}{c|cc|cc|cc|cc}
\hline
Error    & \multicolumn{2}{c|}{KL}                  & \multicolumn{2}{c|}{$I_{cl}$}            & \multicolumn{2}{c|}{$I$}                       & \multicolumn{2}{c}{Energy}              \\ \hline
$\beta$  & \multicolumn{1}{c|}{0}     & $\infty$    & \multicolumn{1}{c|}{0}     & $\infty$    & \multicolumn{1}{c|}{0}        & $\infty$       & \multicolumn{1}{c|}{0}     & $\infty$   \\ \hline
Shadows  & \multicolumn{2}{c|}{\multirow{3}{*}{-1}} & \multicolumn{2}{c|}{\multirow{3}{*}{-1}} & \multicolumn{2}{c|}{\multirow{2}{*}{/}} & \multicolumn{2}{c}{\multirow{2}{*}{-2}} \\
POVM-NQS & \multicolumn{2}{c|}{}                    & \multicolumn{2}{c|}{}                    & \multicolumn{2}{c|}{}                          & \multicolumn{2}{c}{}                    \\ \cline{6-9} 
NDO      & \multicolumn{2}{c|}{}                    & \multicolumn{2}{c|}{}                    & \multicolumn{2}{c|}{valley}                    & \multicolumn{1}{c|}{-2}    & -1         \\ \hline
\end{tabular}
\endgroup

\caption{Scaling exponents for KL, classical infidelity, infidelity, and energy error for TFIM under different $\beta$ with different methods. Valley means $-1$ for $\beta\to 0, \infty$ while decreasing to $-2$ for the intermediate regime.}
\label{tab:exponents}
\end{table}

The previous section discussed the asymptotic quantum shot complexity for both KL and classical infidelity.
Now, our focus shifts to examining the error scaling on the energy and quantum infidelity, intending to provide a theoretical explanation for the observed behavior. 
To this end, we consider a simple phenomenological model of errors occurring in NQS reconstructions. 
After the training procedure, we assume that the NQS does not perfectly encode the target state, but has a small error denoted as $\delta$.
We then derive asymptotic expressions to identify how this error affects various error metrics.
The findings are summarized in \cref{tab:exponents}.

For POVM-NQS, which directly encodes the probability distribution of POVM outcomes, we make the assumption that
\begin{equation}
    q^\theta(\sigma)=p(\sigma) + \delta\Delta(\sigma)+o(\delta^2),
\end{equation}
where $\sum_\sigma \Delta(\sigma)=0$.
This implies that the total variation distance $\mathrm{TV}$, defined as $\sum_\sigma |p(\sigma) - q^\theta(\sigma)|/2$ is of order $\delta$. 
Building upon the theory of classical distributions learning, we understand that the sample complexity scales as $\mathrm{TV}^{-2}\sim \mathrm{\delta}^{-2}$ \cite{canonne2020short}. 
Since the energy can be expressed as an expectation over the POVM distribution $\sum_{\sigma}p(\sigma)H_{\sigma}$, with $H_{\sigma} = \tr(P_{\sigma'}H)T^{-1}_{\sigma'\sigma}$ \cite{schmale2022efficient}, the error in energy is also of order $\delta$.
As a result, the energy error exhibits an exponent of $-2$. 

For NDO, which directly encodes the density matrix elements, we consider:
\begin{equation}
    \rho_\theta = \rho + \delta \Delta + o(\delta^2),
\end{equation}
where $\tr\Delta=0$.
The trace distance $\mathrm{TD}$, defined as $\|\rho^\theta-\rho\|_1/2$ will therefore be of order $\delta$. 
According to the theory of quantum state learning, the sample complexity scales as $\mathrm{TD}^{-2}\sim \delta^{-2}$ \cite{haah2016sample}.
The error in energy reads $|\tr(\rho_\theta H)-\tr(\rho H)| = \delta |\tr(\Delta H)|+o(\delta^2)$, which is in general of order $\delta$. 
However, when the state is pure and an eigenstate of $H$, we can let $\rho = \ket{\psi}\bra{\psi}$ and $\rho_\theta = (\ket{\psi}+\delta \ket{\Delta})(\bra{\psi}+\delta \bra{\Delta})$, where $\bra{\Delta}\ket{\psi}=0$. 
Then $\tr(\Delta H) = 2\mathrm{Re} \matrixelement{\Delta}{H}{\psi}=0$, since $\ket{\psi}$ is an eigenstate. 
Thus the error in energy is of order $\delta^2$ when $\rho$ is a pure eigenstate, and of order $\delta^2$ otherwise.
This distinction arises from the cancellation of terms of order $\delta$ when the state is pure and an eigenstate of the Hamiltonian. 
However, such cancellation doesn't occur for mixed states or generic observables, leading to the observed scaling behavior: the energy has an exponent of $-1$ for pure states, deteriorating to $-2$ as mixedness increases.

Furthermore, in \cref{app:theory}, we show that KL is of order $\delta^2$ and establish a general proposition that the trace distance can be upper bounded by the square root of the KL.
This finding is consistent with the exponent of $-1$ for KL.

The behavior of quantum infidelity presents a more complex scenario.
According to the theory of quantum state tomography \cite{flammia2012quantum, haah2016sample, yuen2023improved}, infidelity should exhibit an exponent of $-1$, which agrees with the NDO simulations when mixedness is small or large. 
However, in the regime of intermediate mixedness, we observe a degradation of the infidelity exponent, forming a so-called \textit{valley}.
We explain this phenomenon in \cref{app:valley} due to the misalignment between KL and infidelity, and we can reproduce a qualitatively consistent valley with random reconstruction error $\Delta$ in numerical simulations. 
The remaining quantitative discrepancies might arise from complicated interactions between the NN structure, training heuristics, and properties of the target states. 

For POVM-NQS, we observe that a simple switch from $I(\rho, \rho_\theta)$ to $I(\rho_\theta, \rho)$ significantly alters the behavior, indicating the presence of several negative eigenvalues in $\rho_\theta$. 
This suggests that the observed behavior of quantum infidelity is primarily caused by the unphysical nature of POVM-NQS reconstruction and is presented here only for completeness.
We note that the same reason caused the anomalous behavior of quantum infidelity in the right top panel of \cref{fig:scaling_ising}.

\section{Conclusions}

In this paper, we systematically study the sample complexity of NQS mixed-state reconstruction and compare different NQS encodings, including NDO and POVM-NQS. 
To achieve accurate reconstruction, we introduce a strategy to systematically suppress the noise introduced by mini-batching based on Control-Variates.
We provide theoretical arguments and numerical proof that this strategy leads to significantly better accuracy of reconstruction algorithms and has no trade-offs.
Even though we only discuss the case of mixed-state reconstruction, it can also be applied to any scheme based on NQS.
We also open-sourced a high-quality implementation in the quantum state reconstruction (QSR) driver of NetKet \cite{netket2:2019, netket3:2022}.

We then present extensive numerical simulations for the finite-temperature TFIM, which is a prototypical example of realistic scenarios that experimentalists would encounter in quantum simulation experiments on NISQ devices. 
We find that NDO offers a quadratic advantage over POVM-NQS and classical shadows in the asymptotic sample complexity when the state is pure or almost pure.
This advantage deteriorates and eventually vanishes when the target state becomes more mixed.
On the other hand, POVM-NQS treats states of various mixedness on an equal footing and does not have such an advantage at all, regardless of the state's mixedness. 
Therefore, NDO is a more efficient tool for state reconstruction for slightly mixed states.

Our results establish asymptotic sample complexity as an important performance indicator for designing NQS architectures and showcase the advantages of enforcing physical constraints at the level of the NN architecture.
We also provides a first comparison of the performance of the NDO and POVM-NQS encodings for mixed-states, which has otherwise not been investigated and might be of interest for developing variational methods to simulate finite-temperature and/or open quantum systems.

Finally, we note that in this work we focus on the error dependence of sample complexity, and the system size dependence for general mixed states using different ansazte remains to be elucidated.
A more systematic study of this issue is computationally challenging as we remarked in \cref{app:numeric}, and we leave this for future works.

\begin{acknowledgments}
We thank J. Carrasquilla for insightful discussions.
We also thank the anonymous reviewers for constructive suggestions.
We acknowledge the Tsinghua Astrophysics High-Performance Computing platform and SCITAS (EPFL) for providing computational and data storage resources. 
\end{acknowledgments}

\bibliographystyle{quantum}
\bibliography{apssamp}

\providecommand{\noopsort}[1]{}\providecommand{\singleletter}[1]{#1}%
\begin{thebibliography}{10}

\bibitem{nielsen2010quantum}
Michael~A Nielsen and Isaac~L Chuang.
\newblock ``Quantum computation and quantum information''.
\newblock
  \href{https://dx.doi.org/https://doi.org/10.1017/CBO9780511976667}{Cambridge
  University Press}. ~(2010).

\bibitem{georgescu2014quantum}
Iulia~M Georgescu, Sahel Ashhab, and Franco Nori.
\newblock ``Quantum simulation''.
\newblock
  \href{https://dx.doi.org/https://doi.org/10.1103/RevModPhys.86.153}{Reviews
  of Modern Physics {\bf 86}, 153}~(2014).

\bibitem{gisin2007quantum}
Nicolas Gisin and Rob Thew.
\newblock ``Quantum communication''.
\newblock
  \href{https://dx.doi.org/https://doi.org/10.1038/nphoton.2007.22}{Nature
  Photonics {\bf 1}, 165--171}~(2007).

\bibitem{gisin2002quantum}
Nicolas Gisin, Gr{\'e}goire Ribordy, Wolfgang Tittel, and Hugo Zbinden.
\newblock ``Quantum cryptography''.
\newblock
  \href{https://dx.doi.org/https://doi.org/10.1103/revmodphys.74.145}{Reviews
  of Modern Physics {\bf 74}, 145}~(2002).

\bibitem{biamonte2017quantum}
Jacob Biamonte, Peter Wittek, Nicola Pancotti, Patrick Rebentrost, Nathan
  Wiebe, and Seth Lloyd.
\newblock ``Quantum machine learning''.
\newblock \href{https://dx.doi.org/https://doi.org/10.1038/nature23474}{Nature
  {\bf 549}, 195--202}~(2017).

\bibitem{preskill2018quantum}
John Preskill.
\newblock ``Quantum computing in the nisq era and beyond''.
\newblock
  \href{https://dx.doi.org/https://doi.org/10.22331/q-2018-08-06-79}{Quantum
  {\bf 2}, 79}~(2018).

\bibitem{cai2022quantum}
Zhenyu Cai, Ryan Babbush, Simon~C Benjamin, Suguru Endo, William~J Huggins,
  Ying Li, Jarrod~R McClean, and Thomas~E O’Brien.
\newblock ``Quantum error mitigation''.
\newblock
  \href{https://dx.doi.org/https://doi.org/10.1103/RevModPhys.95.045005}{Reviews
  of Modern Physics {\bf 95}, 045005}~(2023).

\bibitem{mcclean2016theory}
Jarrod~R McClean, Jonathan Romero, Ryan Babbush, and Al{\'a}n Aspuru-Guzik.
\newblock ``The theory of variational hybrid quantum-classical algorithms''.
\newblock
  \href{https://dx.doi.org/https://doi.org/10.1088/1367-2630/18/2/023023}{New
  Journal of Physics {\bf 18}, 023023}~(2016).

\bibitem{carrasco2021theoretical}
Jose Carrasco, Andreas Elben, Christian Kokail, Barbara Kraus, and Peter
  Zoller.
\newblock ``Theoretical and experimental perspectives of quantum
  verification''.
\newblock
  \href{https://dx.doi.org/https://doi.org/10.1103/prxquantum.2.010102}{PRX
  Quantum {\bf 2}, 010102}~(2021).

\bibitem{paris2004quantum}
Matteo Paris and Jaroslav Rehacek.
\newblock ``Quantum state estimation''.
\newblock \href{https://dx.doi.org/https://doi.org/10.1007/b98673}{Volume 649}.
\newblock Springer Science \& Business Media. ~(2004).

\bibitem{lvovsky2004iterative}
Alexander~I Lvovsky.
\newblock ``Iterative maximum-likelihood reconstruction in quantum homodyne
  tomography''.
\newblock
  \href{https://dx.doi.org/https://doi.org/10.1088/1464-4266/6/6/014}{Journal
  of Optics B: Quantum and Semiclassical Optics {\bf 6}, S556}~(2004).

\bibitem{vrehavcek2001iterative}
J~{\v{R}}eh{\'a}{\v{c}}ek, Z~Hradil, and M~Je{\v{z}}ek.
\newblock ``Iterative algorithm for reconstruction of entangled states''.
\newblock
  \href{https://dx.doi.org/https://doi.org/10.1103/physreva.63.040303}{Physical
  Review A {\bf 63}, 040303}~(2001).

\bibitem{flammia2012quantum}
Steven~T Flammia, David Gross, Yi-Kai Liu, and Jens Eisert.
\newblock ``Quantum tomography via compressed sensing: error bounds, sample
  complexity and efficient estimators''.
\newblock
  \href{https://dx.doi.org/https://doi.org/10.1088/1367-2630/14/9/095022}{New
  Journal of Physics {\bf 14}, 095022}~(2012).

\bibitem{haah2016sample}
Jeongwan Haah, Aram~W Harrow, Zhengfeng Ji, Xiaodi Wu, and Nengkun Yu.
\newblock ``Sample-optimal tomography of quantum states''.
\newblock In Proceedings of the forty-eighth annual ACM symposium on Theory of
  Computing.
\newblock
  \href{https://dx.doi.org/https://doi.org/10.1145/2897518.2897585}{Pages
  913--925}.
\newblock ~(2016).

\bibitem{yuen2023improved}
Henry Yuen.
\newblock ``An improved sample complexity lower bound for (fidelity) quantum
  state tomography''.
\newblock
  \href{https://dx.doi.org/https://doi.org/10.22331/q-2023-01-03-890}{Quantum
  {\bf 7}, 890}~(2023).

\bibitem{aaronson2018shadow}
Scott Aaronson.
\newblock ``Shadow tomography of quantum states''.
\newblock In Proceedings of the 50th annual ACM SIGACT symposium on theory of
  computing.
\newblock
  \href{https://dx.doi.org/https://doi.org/10.1145/3188745.3188802}{Pages
  325--338}.
\newblock ~(2018).

\bibitem{huang2020predicting}
Hsin-Yuan Huang, Richard Kueng, and John Preskill.
\newblock ``Predicting many properties of a quantum system from very few
  measurements''.
\newblock
  \href{https://dx.doi.org/https://doi.org/10.1038/s41567-020-0932-7}{Nature
  Physics {\bf 16}, 1050--1057}~(2020).

\bibitem{hadfield2022measurements}
Charles Hadfield, Sergey Bravyi, Rudy Raymond, and Antonio Mezzacapo.
\newblock ``Measurements of quantum hamiltonians with locally-biased classical
  shadows''.
\newblock
  \href{https://dx.doi.org/https://doi.org/10.1007/s00220-022-04343-8}{Communications
  in Mathematical Physics {\bf 391}, 951--967}~(2022).

\bibitem{elben2023randomized}
Andreas Elben, Steven~T Flammia, Hsin-Yuan Huang, Richard Kueng, John Preskill,
  Beno{\^\i}t Vermersch, and Peter Zoller.
\newblock ``The randomized measurement toolbox''.
\newblock
  \href{https://dx.doi.org/https://doi.org/10.1038/s42254-022-00535-2}{Nature
  Reviews Physics {\bf 5}, 9--24}~(2023).

\bibitem{poulin2011quantum}
David Poulin, Angie Qarry, Rolando Somma, and Frank Verstraete.
\newblock ``Quantum simulation of time-dependent hamiltonians and the
  convenient illusion of hilbert space''.
\newblock
  \href{https://dx.doi.org/https://doi.org/10.1103/physrevlett.106.170501}{Physical
  Review Letters {\bf 106}, 170501}~(2011).

\bibitem{brandao2021models}
Fernando~GSL Brand{\~a}o, Wissam Chemissany, Nicholas Hunter-Jones, Richard
  Kueng, and John Preskill.
\newblock ``Models of quantum complexity growth''.
\newblock
  \href{https://dx.doi.org/https://doi.org/10.1103/prxquantum.2.030316}{PRX
  Quantum {\bf 2}, 030316}~(2021).

\bibitem{zhao2023learning}
Haimeng Zhao, Laura Lewis, Ishaan Kannan, Yihui Quek, Hsin-Yuan Huang, and
  Matthias~C Caro.
\newblock ``Learning quantum states and unitaries of bounded gate
  complexity''~(2023).
\newblock  \href{http://arxiv.org/abs/2310.19882}{arXiv:2310.19882}.

\bibitem{baumgratz2013scalable}
Tillmann Baumgratz, Alexander N{\"u}{\ss}eler, Marcus Cramer, and Martin~B
  Plenio.
\newblock ``A scalable maximum likelihood method for quantum state
  tomography''.
\newblock
  \href{https://dx.doi.org/https://doi.org/10.1088/1367-2630/15/12/125004}{New
  Journal of Physics {\bf 15}, 125004}~(2013).

\bibitem{lanyon2017efficient}
BP~Lanyon, C~Maier, Milan Holz{\"a}pfel, Tillmann Baumgratz, C~Hempel,
  P~Jurcevic, Ish Dhand, AS~Buyskikh, AJ~Daley, Marcus Cramer, et~al.
\newblock ``Efficient tomography of a quantum many-body system''.
\newblock \href{https://dx.doi.org/https://doi.org/10.1038/nphys4244}{Nature
  Physics {\bf 13}, 1158--1162}~(2017).

\bibitem{guo2024scalable}
Yuchen Guo and Shuo Yang.
\newblock ``Scalable quantum state tomography with locally purified density
  operators and local measurements''~(2024).
\newblock  \href{http://arxiv.org/abs/2307.16381}{arXiv:2307.16381}.

\bibitem{carleo2017solving}
Giuseppe Carleo and Matthias Troyer.
\newblock ``Solving the quantum many-body problem with artificial neural
  networks''.
\newblock
  \href{https://dx.doi.org/https://doi.org/10.1126/science.aag2302}{Science
  {\bf 355}, 602--606}~(2017).

\bibitem{torlai2018neural}
Giacomo Torlai, Guglielmo Mazzola, Juan Carrasquilla, Matthias Troyer, Roger
  Melko, and Giuseppe Carleo.
\newblock ``Neural-network quantum state tomography''.
\newblock
  \href{https://dx.doi.org/https://doi.org/10.1038/s41567-018-0048-5}{Nature
  Physics {\bf 14}, 447--450}~(2018).

\bibitem{carrasquilla2019reconstructing}
Juan Carrasquilla, Giacomo Torlai, Roger~G Melko, and Leandro Aolita.
\newblock ``Reconstructing quantum states with generative models''.
\newblock
  \href{https://dx.doi.org/https://doi.org/10.1038/s42256-019-0028-1}{Nature
  Machine Intelligence {\bf 1}, 155--161}~(2019).

\bibitem{lohani2021experimental}
Sanjaya Lohani, Thomas~A Searles, Brian~T Kirby, and Ryan~T Glasser.
\newblock ``On the experimental feasibility of quantum state reconstruction via
  machine learning''.
\newblock
  \href{https://dx.doi.org/https://doi.org/10.1109/tqe.2021.3106958}{IEEE
  Transactions on Quantum Engineering {\bf 2}, 1--10}~(2021).

\bibitem{schmale2022efficient}
Tobias Schmale, Moritz Reh, and Martin G{\"a}rttner.
\newblock ``Efficient quantum state tomography with convolutional neural
  networks''.
\newblock
  \href{https://dx.doi.org/https://doi.org/10.1038/s41534-022-00621-4}{npj
  Quantum Information {\bf 8}, 115}~(2022).

\bibitem{cha2021attention}
Peter Cha, Paul Ginsparg, Felix Wu, Juan Carrasquilla, Peter~L McMahon, and
  Eun-Ah Kim.
\newblock ``Attention-based quantum tomography''.
\newblock
  \href{https://dx.doi.org/https://doi.org/10.1088/2632-2153/ac362b}{Machine
  Learning: Science and Technology {\bf 3}, 01LT01}~(2021).

\bibitem{torlai2018latent}
Giacomo Torlai and Roger~G Melko.
\newblock ``Latent space purification via neural density operators''.
\newblock
  \href{https://dx.doi.org/https://doi.org/10.1103/physrevlett.120.240503}{Physical
  Review Letters {\bf 120}, 240503}~(2018).

\bibitem{goodfellow2016deep}
Ian Goodfellow, Yoshua Bengio, and Aaron Courville.
\newblock ``Deep learning''.
\newblock MIT Press. ~(2016).
\newblock  url:~\url{http://www.deeplearningbook.org}.

\bibitem{torlai2020precise}
Giacomo Torlai, Guglielmo Mazzola, Giuseppe Carleo, and Antonio Mezzacapo.
\newblock ``Precise measurement of quantum observables with neural-network
  estimators''.
\newblock
  \href{https://dx.doi.org/https://doi.org/10.1103/physrevresearch.2.022060}{Physical
  Review Research {\bf 2}, 022060}~(2020).

\bibitem{Melkani2020PRAEigenstateExtraction}
Abhijeet Melkani, Clemens Gneiting, and Franco Nori.
\newblock ``Eigenstate extraction with neural-network tomography''.
\newblock \href{https://dx.doi.org/10.1103/PhysRevA.102.022412}{Physical Review
  A {\bf 102}, 022412}~(2020).

\bibitem{vicentini2022positive}
Filippo Vicentini, Riccardo Rossi, and Giuseppe Carleo.
\newblock ``Positive-definite parametrization of mixed quantum states with deep
  neural networks''~(2022).
\newblock  \href{http://arxiv.org/abs/2206.13488}{arXiv:2206.13488}.

\bibitem{Neugebauer20PRAPovmVsNQS}
Marcel Neugebauer, Laurin Fischer, Alexander J\"ager, Stefanie Czischek, Selim
  Jochim, Matthias Weidem\"uller, and Martin G\"arttner.
\newblock ``Neural-network quantum state tomography in a two-qubit
  experiment''.
\newblock \href{https://dx.doi.org/10.1103/PhysRevA.102.042604}{Physical Review
  A {\bf 102}, 042604}~(2020).

\bibitem{sehayek2019learnability}
Dan Sehayek, Anna Golubeva, Michael~S Albergo, Bohdan Kulchytskyy, Giacomo
  Torlai, and Roger~G Melko.
\newblock ``Learnability scaling of quantum states: Restricted boltzmann
  machines''.
\newblock
  \href{https://dx.doi.org/https://doi.org/10.1103/physrevb.100.195125}{Physical
  Review B {\bf 100}, 195125}~(2019).

\bibitem{kandala2017hardware}
Abhinav Kandala, Antonio Mezzacapo, Kristan Temme, Maika Takita, Markus Brink,
  Jerry~M Chow, and Jay~M Gambetta.
\newblock ``Hardware-efficient variational quantum eigensolver for small
  molecules and quantum magnets''.
\newblock \href{https://dx.doi.org/https://doi.org/10.1038/nature23879}{Nature
  {\bf 549}, 242--246}~(2017).

\bibitem{lukens2021bayesian}
Joseph~M Lukens, Kody~JH Law, and Ryan~S Bennink.
\newblock ``A bayesian analysis of classical shadows''.
\newblock
  \href{https://dx.doi.org/https://doi.org/10.1038/s41534-021-00447-6}{npj
  Quantum Information {\bf 7}, 113}~(2021).

\bibitem{iouchtchenko2023neural}
Dmitri Iouchtchenko, J{\'e}r{\^o}me~F Gonthier, Alejandro Perdomo-Ortiz, and
  Roger~G Melko.
\newblock ``Neural network enhanced measurement efficiency for molecular
  groundstates''.
\newblock
  \href{https://dx.doi.org/https://doi.org/10.1088/2632-2153/acb4df}{Machine
  Learning: Science and Technology {\bf 4}, 015016}~(2023).

\bibitem{carrasquilla2021probabilistic}
Juan Carrasquilla, Di~Luo, Felipe P{\'e}rez, Ashley Milsted, Bryan~K Clark,
  Maksims Volkovs, and Leandro Aolita.
\newblock ``Probabilistic simulation of quantum circuits using a deep-learning
  architecture''.
\newblock
  \href{https://dx.doi.org/https://doi.org/10.1103/physreva.104.032610}{Physical
  Review A {\bf 104}, 032610}~(2021).

\bibitem{becca2017quantum}
Federico Becca and Sandro Sorella.
\newblock ``Quantum monte carlo approaches for correlated systems''.
\newblock
  \href{https://dx.doi.org/https://doi.org/10.1017/9781316417041}{Cambridge
  University Press}. ~(2017).

\bibitem{sinibaldi2023unbiasing}
Alessandro Sinibaldi, Clemens Giuliani, Giuseppe Carleo, and Filippo Vicentini.
\newblock ``Unbiasing time-dependent variational monte carlo by projected
  quantum evolution''.
\newblock \href{https://dx.doi.org/10.22331/q-2023-10-10-1131}{Quantum {\bf 7},
  1131}~(2023).

\bibitem{boyle1977options}
Phelim~P Boyle.
\newblock ``Options: A monte carlo approach''.
\newblock
  \href{https://dx.doi.org/https://doi.org/10.1016/0304-405X(77)90005-8}{Journal
  of Financial Economics {\bf 4}, 323--338}~(1977).

\bibitem{wu_solving_2019}
Dian Wu, Lei Wang, and Pan Zhang.
\newblock ``Solving {Statistical} {Mechanics} {Using} {Variational}
  {Autoregressive} {Networks}''.
\newblock \href{https://dx.doi.org/10.1103/PhysRevLett.122.080602}{Physical
  Review Letters {\bf 122}, 080602}~(2019).

\bibitem{sharir_deep_2020}
Or~Sharir, Yoav Levine, Noam Wies, Giuseppe Carleo, and Amnon Shashua.
\newblock ``Deep {Autoregressive} {Models} for the {Efficient} {Variational}
  {Simulation} of {Many}-{Body} {Quantum} {Systems}''.
\newblock \href{https://dx.doi.org/10.1103/PhysRevLett.124.020503}{Physical
  Review Letters {\bf 124}, 020503}~(2020).

\bibitem{hibat2020recurrent}
Mohamed Hibat-Allah, Martin Ganahl, Lauren~E Hayward, Roger~G Melko, and Juan
  Carrasquilla.
\newblock ``Recurrent neural network wave functions''.
\newblock
  \href{https://dx.doi.org/https://doi.org/10.1103/physrevresearch.2.023358}{Physical
  Review Research {\bf 2}, 023358}~(2020).

\bibitem{johnson2013accelerating}
Rie Johnson and Tong Zhang.
\newblock ``Accelerating stochastic gradient descent using predictive variance
  reduction''.
\newblock Advances in Neural Information Processing Systems{\bf 26}~(2013).
\newblock
  url:~\url{https://proceedings.neurips.cc/paper_files/paper/2013/file/ac1dd209cbcc5e5d1c6e28598e8cbbe8-Paper.pdf}.

\bibitem{mohamed2020monte}
Shakir Mohamed, Mihaela Rosca, Michael Figurnov, and Andriy Mnih.
\newblock ``Monte carlo gradient estimation in machine learning''.
\newblock Journal of Machine Learning Research {\bf 21}, 1--62~(2020).
\newblock  url:~\url{http://jmlr.org/papers/v21/19-346.html}.

\bibitem{netket2:2019}
Giuseppe Carleo, Kenny Choo, Damian Hofmann, James E.~T. Smith, Tom Westerhout,
  Fabien Alet, Emily~J. Davis, Stavros Efthymiou, Ivan Glasser, Sheng-Hsuan
  Lin, Marta Mauri, Guglielmo Mazzola, Christian~B. Mendl, Evert van
  Nieuwenburg, Ossian O'Reilly, Hugo Th{\'e}veniaut, Giacomo Torlai, Filippo
  Vicentini, and Alexander Wietek.
\newblock ``Netket: A machine learning toolkit for many-body quantum systems''.
\newblock \href{https://dx.doi.org/10.1016/j.softx.2019.100311}{SoftwareXPage
  100311}~(2019).

\bibitem{netket3:2022}
Filippo Vicentini, Damian Hofmann, Attila Szabó, Dian Wu, Christopher Roth,
  Clemens Giuliani, Gabriel Pescia, Jannes Nys, Vladimir Vargas-Calderón,
  Nikita Astrakhantsev, and Giuseppe Carleo.
\newblock ``{NetKet 3: Machine Learning Toolbox for Many-Body Quantum
  Systems}''.
\newblock \href{https://dx.doi.org/10.21468/SciPostPhysCodeb.7}{SciPost Physics
  CodebasesPage~7}~(2022).

\bibitem{mpi4jax:2021}
Dion Häfner and Filippo Vicentini.
\newblock ``mpi4jax: Zero-copy mpi communication of jax arrays''.
\newblock \href{https://dx.doi.org/10.21105/joss.03419}{Journal of Open Source
  Software {\bf 6}, 3419}~(2021).

\bibitem{arute2019quantum}
Frank Arute, Kunal Arya, Ryan Babbush, Dave Bacon, Joseph~C Bardin, Rami
  Barends, Rupak Biswas, Sergio Boixo, Fernando~GSL Brandao, David~A Buell,
  et~al.
\newblock ``Quantum supremacy using a programmable superconducting processor''.
\newblock
  \href{https://dx.doi.org/https://doi.org/10.1038/s41586-019-1666-5}{Nature
  {\bf 574}, 505--510}~(2019).

\bibitem{endres2016atom}
Manuel Endres, Hannes Bernien, Alexander Keesling, Harry Levine, Eric~R
  Anschuetz, Alexandre Krajenbrink, Crystal Senko, Vladan Vuletic, Markus
  Greiner, and Mikhail~D Lukin.
\newblock ``Atom-by-atom assembly of defect-free one-dimensional cold atom
  arrays''.
\newblock
  \href{https://dx.doi.org/https://doi.org/10.1126/science.aah3752}{Science
  {\bf 354}, 1024--1027}~(2016).

\bibitem{canonne2020short}
Cl{\'e}ment~L Canonne.
\newblock ``A short note on learning discrete distributions''~(2020).
\newblock  \href{http://arxiv.org/abs/2002.11457}{arXiv:2002.11457}.

\bibitem{vicentini2019variational}
Filippo Vicentini, Alberto Biella, Nicolas Regnault, and Cristiano Ciuti.
\newblock ``Variational neural-network ansatz for steady states in open quantum
  systems''.
\newblock
  \href{https://dx.doi.org/https://doi.org/10.1103/physrevlett.122.250503}{Physical
  Review Letters {\bf 122}, 250503}~(2019).

\bibitem{kingma2014adam}
Diederik~P Kingma and Jimmy Ba.
\newblock ``Adam: A method for stochastic optimization''.
\newblock In International Conference on Learning Representations.
\newblock San Diega, CA, USA~(2015).
\newblock  \href{http://arxiv.org/abs/1412.6980}{arXiv:1412.6980}.

\bibitem{jax2018github}
James Bradbury, Roy Frostig, Peter Hawkins, Matthew~James Johnson, Chris Leary,
  Dougal Maclaurin, George Necula, Adam Paszke, Jake Vander{P}las, Skye
  Wanderman-{M}ilne, and Qiao Zhang.
\newblock ``{JAX}: composable transformations of {P}ython+{N}um{P}y
  programs''~(2018).
\newblock \url{http://github.com/google/jax}.

\bibitem{cover1999elements}
Thomas~M Cover and Joy~A. Thomas.
\newblock ``Elements of information theory''.
\newblock \href{https://dx.doi.org/https://doi.org/10.1002/047174882x}{John
  Wiley \& Sons}. ~(1999).

\bibitem{zhao2023non}
Haimeng Zhao.
\newblock ``Non-iid quantum federated learning with one-shot communication
  complexity''.
\newblock
  \href{https://dx.doi.org/https://doi.org/10.1007/s42484-022-00091-z}{Quantum
  Machine Intelligence {\bf 5}, 3}~(2023).

\bibitem{liu2022cps}
Junyi Liu, Yifu Tang, Haimeng Zhao, Xieheng Wang, Fangyu Li, and Jingyi Zhang.
\newblock ``Cps attack detection under limited local information in cyber
  security: An ensemble multi-node multi-class classification approach''.
\newblock \href{https://dx.doi.org/10.1145/3585520}{ACM Transactions on Sensor
  Networks {\bf 20}, 1–27}~(2024).

\end{thebibliography}

\appendix

\section{Numerical Details} \label{app:numeric}
Here we list the numerical details of our studies. 
The NDO used in the simulations of an $N$-qubit system is a restricted Boltzmann machine with one layer of $N$ hidden neurons and $N$ ancillas \cite{vicentini2019variational}.
The POVM-NQS is an autoregressive dense NN with 2 layers of 10 neurons \cite{schmale2022efficient}. 
All training are conducted via the Adam optimizer \cite{kingma2014adam} with learning rate $10^{-3}$, batch size $100$, maximal iteration number $10^5$ and CV update frequency $T=50$. 
The training is terminated when the loss value stops decreasing for 2000 iterations.
All code is implemented with NetKet \cite{netket2:2019, netket3:2022, mpi4jax:2021} and JAX \cite{jax2018github}.

We note that the numerical study of the NDO scaling for mixed states is computationally more challenging than the case of pure states, such as Ref.~\cite{iouchtchenko2023neural}.
This is because for pure states, measurements of all the Pauli strings in the Hamiltonian suffice to determine the state, while for mixed states they don't.
Intuitively, when the pure state ansatz is trained to reproduce the probability distributions of all Hamiltonian terms, it gives an energy approximating the true energy, which is the minimal energy. 
Then by the variational principle, the state also approximates the ground state.
In contrast, for mixed states, one has to measure an informationally-complete set of bases (e.g., all the Pauli bases) to minimize the reconstruction error.
This exponentially growing basis-set size makes numerical simulations for larger systems very challenging.
Nevertheless, our theoretical analysis is independent of the system size and agrees with numerical simulations on small systems. 
Also, a relevant open question would be to investigate what is the effect of a truncated basis set on the reconstruction accuracy, which is relevant for experimental implementation. 

\section{Theoretical Details} \label{app:theory}
For NDO, we assume $\rho_\theta = \rho + \delta\Delta + o(\delta^2)$, and we will omit $o(\delta^2)$ in derivations in this appendix.
The KL divergence reads
\begin{equation}
q^\theta_b(\sigma) = \tr((\rho+\delta\Delta)P^b_\sigma) = p_b(\sigma) + \delta\tr(\Delta P^b_\sigma).
\end{equation}
Thus $\mathrm{KL}(p_b\|q^\theta_b) = -\sum_\sigma p_b(\sigma)\log\left(1+\delta\frac{\tr(\Delta P^b_\sigma)}{p_b(\sigma)}\right) = -\delta\tr(\Delta\sum_\sigma P^b_\sigma)=o(\delta^2)$, where we have used $\sum_\sigma P^b_\sigma=I$ and $\tr\Delta=0$. 
Therefore $\mathrm{KL}=\sum_{b=1}^{N_b}\mathrm{KL}(p_b\|q^\theta_b)/N_b=o(\delta^2)$.

In fact, apart from the phenomenological model, we can also derive the quadratic relation between KL and trace distance via the following inequality, which is model-independent.
\begin{proposition}
(Trace distance bounded by KL over all Pauli basis).
For two $n$-qubit states $\rho$ and $\rho'$, let $\mathrm{TD}=\|\rho-\rho'\|_1/2$ be the trace distance and use $\sigma_i\in\{I, X, Y, Z\}$ to denote Pauli operators. 
Define the KL over all Pauli basis as $\mathrm{KL}=\sum_{i_1, \ldots, i_n=0}^{3}\mathrm{KL}(p_{i_1, \ldots, i_n}\|p'_{i_1, \ldots, i_n})$, where $p$ and $p'$ denote the Bernoulli probability distributions given by the corresponding Pauli measurements and states. 
Then we have
\begin{equation}
    \mathrm{TD} \leq \frac{2^n}{\sqrt{2}}\sqrt{\mathrm{KL}}.
\end{equation}
\end{proposition}
To show this, note that all Pauli string operators form a complete basis of the space of Hermitian matrices. 
Therefore, we can decompose $\rho$ as
\begin{equation}
    \rho = \sum_{i_1, \cdots, i_n=0}^3 C_{i_1, \cdots, i_n} \sigma_{i_1}\otimes\cdots\otimes\sigma_{i_n}.
\end{equation}
The coefficients
\begin{equation}
    C_{i_1, \cdots, i_n} = \frac{1}{2^n}\mathrm{tr}(\rho  \sigma_{i_1}\otimes\cdots\otimes\sigma_{i_n})=\frac{1}{2^n}\mathbb{E}_{\sigma\sim p_{i_1, \cdots, i_n}}[\sigma],
\end{equation}
where $p_{i_1, \cdots, i_n}$ is the Bernoulli distribution given by $\rho$ measured in the basis $\sigma_{i_1}\otimes\cdots\otimes\sigma_{i_n}$. 
We use the same notations with primes to denote the corresponding quantities of $\rho'$. 
Then the trace distance can be bounded by the differences in coefficients as
\begin{equation} \label{eqn:bound_td_coeff}
\begin{split}
    \mathrm{TD} &\leq \frac{1}{2} \sum_{i_1, \cdots, i_n=0}^3 |C_{i_1, \cdots, i_n}-C'_{i_1, \cdots, i_n}| \|\sigma_{i_1}\otimes\cdots\otimes\sigma_{i_n}\|_1 \\
    &=\frac{1}{2} \sum_{i_1, \cdots, i_n=0}^3 |C_{i_1, \cdots, i_n}-C'_{i_1, \cdots, i_n}|\cdot 2^n.
\end{split}
\end{equation}
On the other hand, the differences in coefficients can be bounded as
\begin{equation} \label{eqn:bound_coeff_kl}
\begin{split}
    &|(C_{i_1, \cdots, i_n}-C'_{i_1, \cdots, i_n})| \\
    &= \frac{1}{2^n} \left|\mathbb{E}_{\sigma\sim p_{i_1, \cdots, i_n}}[\sigma] - \mathbb{E}_{\sigma\sim p'_{i_1, \cdots, i_n}}[\sigma] \right| \\
    &\le \frac{1}{2^n} \sum_{\sigma\in\{\pm 1\}} |p_{i_1, \cdots, i_n}(\sigma) - p'_{i_1, \cdots, i_n}(\sigma)| \\
    &\le \frac{2}{2^n} \sqrt{\frac{1}{2} \mathrm{KL}(p_{i_1, \cdots, i_n} \| p'_{i_1, \cdots, i_n})},
\end{split}
\end{equation}
where the last inequality follows from Pinsker's inequality \cite{cover1999elements}.
Therefore, we arrive at
\begin{equation}
\begin{split}
    \mathrm{TD}(\rho, \rho') &\le \sum_{i_1, \cdots, i_n=0}^3 \sqrt{\frac{1}{2} \mathrm{KL}(p_{i_1, \cdots, i_n} \| p'_{i_1, \cdots, i_n})} \\
    &\le \frac{2^n}{\sqrt{2}} \sqrt{\mathrm{KL}},
\end{split}
\end{equation}
where we have used the mean inequality.
This proposition can also serve as a guarantee for training quantum density estimators in the context of quantum federated learning \cite{zhao2023non, liu2022cps}.

\section{The Valley Phenomenon}
\label{app:valley}

\begin{figure}
    \centering
    \includegraphics[width=\linewidth]{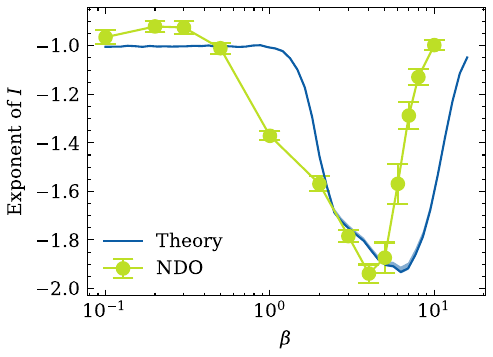}
    \caption{The valley phenomenon in scaling exponents of infidelity for different $\beta$. The exponents observed in NDO simulations are plotted in orange, while the ones given by random perturbations are plotted in blue. Error bars and the shaded region indicate the standard deviation over 100 random instances.}
    \label{fig:valley}
\end{figure}

Here we aim to provide an explanation of the valley phenomenon observed in NDO: the exponent of infidelity is $-1$ for $\beta\to 0, \infty$, while decreasing to $-2$ for intermediate $\beta$.

We start from our phenomenological model and try to find a relationship between KL and infidelity.
Since the behavior of KL is well understood, such a relationship would give us insights into how infidelity behaves.
We consider the error matrix $\Delta$ to be drawn randomly as $\Delta = AA^\dagger/\tr(AA^\dagger)$, where each entry of $A$ follows the complex standard Gaussian distribution.
Then we calculate the perturbed state $\rho_\theta = \rho + \delta \Delta$, and normalize it again by dividing its trace.

For a given $\beta$ and the corresponding target state $\rho$, we randomly generate 100 such perturbed states for different choices of $\delta \in [10^{-3.5}, 10^{-2.5}]$, and calculate the KL and infidelity $I$ against the target state.
We find that the resulting $(\mathrm{KL}, I)$ pairs fall on a straight line in log-log scale, with $r$-squared greater than 0.99.
The slope $\alpha=\alpha(\beta)$ depends on $\beta$, and gives an effective power law relationship between KL and $I$: $I \propto \mathrm{KL}^{\alpha}$.
This means that the way we quantify the reconstruction error impacts the scaling exponents we observe. 
In particular, such misalignment between KL and infidelity leads to a $\beta$-dependent difference of $\alpha(\beta)$ in exponents.

Now we assume that KL has an exponent of $-1$, which is theoretically and numerically validated.
Hence the sample complexity is proportional to $\mathrm{KL}^{-1} \propto I^{-1/\alpha(\beta)}$. 
In \cref{fig:valley}, we plot the simulated exponents $-1/\alpha(\beta)$ in blue, with the standard deviation indicated by the shaded region.
We find a valley pattern that is qualitatively consistent with what we observe in NDO simulations, which confirms our theoretical explanation.
The rest quantitative differences might be a complicated result of the NN design, generalization, training heuristics, and the property of the target states.

\section{Depolarized Molecular Ground-states}
\label{app:LiH}

\begin{figure*}
    \centering
    \includegraphics[width=\linewidth]{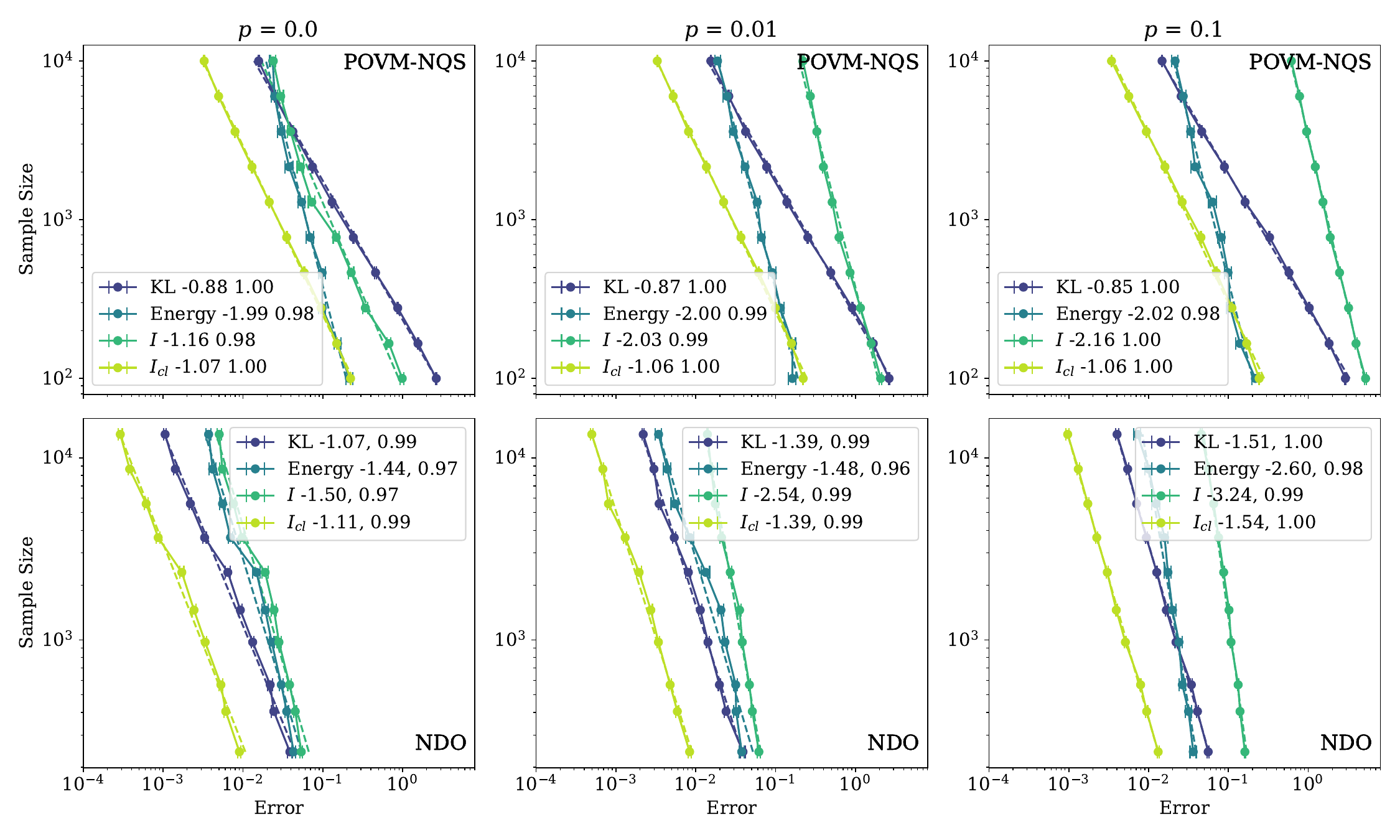}
    \caption{Sample complexity dependence on the reconstruction error for LiH ground-state under different depolarization $p=0, 0.01, 0.1$ using POVM-NQS (upper) and NDO (lower). KL divergence, energy error, infidelity and classical infidelity are used as metrics for reconstruction error. Dashed lines represent the log-log linear regression results, with the slopes and $r^2$ values indicated in the legend. All data points are averaged over 50 random instances.}
    \label{fig:scaling_LiH}
\end{figure*}

In this appendix, we study the sample complexity scaling behavior for LiH ground-state subjected to depolarization noise. 
This scenario emerges in digital simulation or VQEs, where the quantum gates are imperfect and introduce noise that can be modeled as depolarization.
We apply the parity transformation to transform the Fermionic Hamiltonian into a 4-qubit Hamiltonian that can be implemented on quantum computers (details of the transformations are found in the appendix of \cite{torlai2018neural, iouchtchenko2023neural}).
We take its ground state $\ket{\psi}$ and simulate the depolarized states $\rho_p = (1-p)\ket{\psi}\bra{\psi} + pI/2^4$ over $p\in [0, 1]$ as the target states. 
In \cref{fig:scaling_LiH}, we choose three depolarization intensity $p = 0, 0.01, 0.1$ that represents low, medium, and high depolarization regimes, and plot the corresponding results.

We observe that the behavior of POVM-NQS on LiH approximately matches what was seen on the TFIM, while the quantum shot complexity of the NDO is generally lower than that of TFIM by about $0.5$. 
This might arise from the specific NN architecture and training heuristics used here. 
Intuitively, depolarization noise has less structure that can be exploited by NNs than thermal states, leading to a slightly worse scaling in the simulations. 
Nevertheless, an advantage of NDO over POVM-NQS can still be observed at small mixedness, while disappearing when decoherence leads to very mixed states, showing a consistent picture as TFIM.

\end{document}